\documentclass[a4paper,11pt,showkeys,nofootinbib,longbibliography,superscriptaddress]{revtex4-1}
\usepackage{hyperref}  
\usepackage{amsmath, bm}  
\usepackage{graphicx}
\usepackage{url}

\begin{document}

\title{
Comparison of the UNSCEAR isodose maps for annual external exposure in Fukushima with those obtained based on the airborne monitoring surveys}
\author{Ryugo Hayano}
\affiliation{Department of Physics, The University of Tokyo, Tokyo 113-0033, Japan
}
\author{Makoto Miyazaki}
\affiliation{Department of Radiation Health Management, Fukushima Medical University, Fukushima 960-1295, Japan
}

\begin{abstract}
In 2016, UNSCEAR published an attachment to its Fukushima 2015 White Paper,  entitled ``Development of isodose maps representing annual external exposure in Japan as a function of time,'' in which  the committee presented 
annual additional 1 mSv effective dose ab extra isodose lines for 1, 3, 5, 10, 30, 50 years after the accident, based on the  soil deposition data of radionuclides within 100 km from FDNPP.
Meanwhile, the median of the ratio, $c$, between the external effective dose rates and the ambient dose equivalent rates at 1\,m above the ground obtained by the airborne monitoring has been established to be $c\sim 0.15$.  We here compare the UNSCEAR predictions with respect to  estimates based on the airborne monitoring. Although  both methods and data used in the two approaches are different, the resultant contours show relatively good agreement. However, to improve the accuracy of long-term annual effective  isodose lines, feedback from continuous measurements such as airborne monitoring is important.
\end{abstract}
\keywords{Fukushima Dai-ichi Nuclear Power Plant (FDNPP) accident, isodose maps for annual external exposure, UNSCEAR Fukushima 2015 White Paper}
\maketitle

\section{ INTRODUCTION}

In 2016, in Attachment 2 of  UNSCEAR's Fukushima 2015 White Paper (hereafter referred to as Att-2)~\cite{unscear}, as well as in animation~\cite{gif}, UNSCEAR published maps showing  isodose lines for an annual additional effective dose of 1 mSv from external exposure in years 1, 3, 5, 10, 30 and 50 years after the Fukushima Dai-ichi Nuclear Power Plant (FDNPP) accident. 
UNSCEAR estimated isodose lines based on radionuclide soil deposition data obtained from soil samples collected between June 6, 2011 and July 8, 2011, within 100 km from FDNPP. The data were released by the Japanese government in 2011~\cite{JAEA}  and were reproduced by UNSCEAR with geospatial information on a 1 km grid~\cite{c2}.
Att-2 provides a table (Table 2 of \cite{unscear}) of   the conversion coefficients from soil deposition density of radiocaesium to annual effective dose ab extra ($\rm mSv/MBq\, m^{-2}$)  for each year from years 1 to 10, then at every 10 years from years 10 to 50, and 100. Using the soil deposition data and the effective  dose conversion coefficients, it is possible to estimate the annual additional effective dose from external exposure  at each grid point around FDNPP.

Meanwhile, Miyazaki and Hayano~\cite{miyazakihayano1} analyzed data from a large-scale ‘glass-badge’  individual dose  monitoring  (with GIS information of the residents' addresses)  conducted by Date City, Fukushima Prefecture, together with the airborne monitoring data collected periodically by the Japanese government, and found that the ratio $c$ of ambient dose equivalent rate  to the effective dose rate from external exposure is nearly constant between 5 and 51 months after the accident, at $ c \sim 0.15 $~\cite {miyazakihayano1}. Using this relationship, it becomes possible to draw  annual 1 mSv  isodose lines based on the airborne monitoring data.

In this paper, we compare the 1 mSv  annual isodose lines  predicted by UNSCEAR, with those obtained based on the  airborne  monitoring for 3 and 5 years after the accident, and discuss the implications.

\section{MATERIALS AND METHODS}

\subsection{Reproduction of the UNSCEAR isodose lines}
UNSCEAR's  1 mSv annual isodose line maps were created by using 1) the  $ ^ {134} $Cs and $ ^ {137}$ Cs soil deposition density data released by the Japanese government, rebuilt by UNSCEAR on a 1~km mesh in the unit of $\rm MBq\,  m^{-2}$, and 2) dose conversion coefficients from soil  deposition  density to the annual additional effective dose (Table 2 of Att-2). Since the results are shown only as   maps,  we calculated the annual cumulative effective dose at each 1-km grid point, and   reconstructed annual isodose line of 1 mSv, 3 and 5 years after the accident  (by using, respectively, the difference of 2$^{\rm nd}$ and 3$^{\rm rd}$ year cumulative dose coefficients and that of 4$^{\rm th}$ and 5$^{\rm th}$). The agreement was satisfactory but not perfect, due to differences in the interpolation algorithms used for each.

\subsection{Isodose lines based on the airborne monitoring maps}

\begin{figure}
\includegraphics[width=0.7\textwidth]{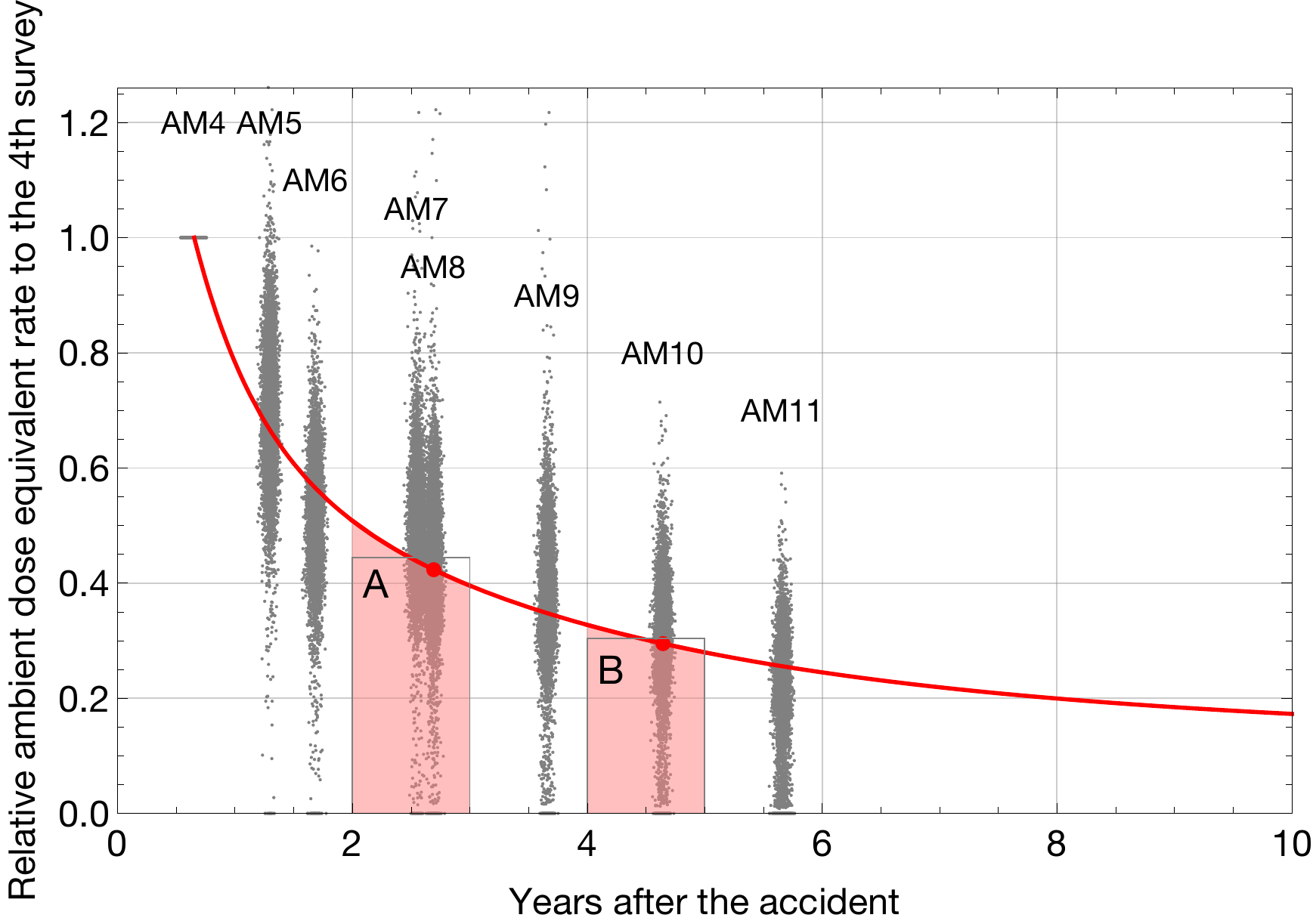}
\caption{\label{fig:decrease} The ratio of the ambient dose equivalent from n$^{\rm th}$ (n=5 to 11) airborne monitoring to that from the 4$^{\rm th}$  for all the airborne monitoring grid points within Fukushima Prefecture are plotted as a function of elapsed time since the accident. The labels AM4...AM11 denote airborne
monitoring 4...11. In order to help alleviate overlaps of points, the data were horizontally jittered. The red curve is a fit to Eq. (1) through all the data points. } 
\end{figure}

 On the other hand, the isodose lines based on the airborne monitoring maps were generated as follows:

For the annual additional effective dose 3 (5)  years after the accident, we used the 8$^{\rm th}$ (10$^{\rm th}$) airborne monitoring data for which the reference date of dose calculation is November 19, 2013 (November 2, 2015), as indicated in Fig.~\ref{fig:decrease}. By multiplying the  ambient dose  equivalent rate $\dot{H}^* (10)$ ($\rm \mu Sv / h$) at each airborne monitoring grid point by the factor $ c = 0.15 $~\cite{miyazakihayano1}, the median value of the external effective dose rate of the population living in the vicinity of the grid point was obtained.

Since the ambient dose equivalent rate 1\,m above the ground $\dot{H}^* (10)$  gradually decreased over time, we cannot simply use the dose rate on the reference date of airborne monitoring as a representative of the whole year. We therefore fitted, as shown in Fig.~\ref{fig:decrease}, the ratios of the ambient dose rate from the  5$^{\rm th}$ through 11$^{\rm th}$ airborne monitoring at each  monitoring grid point to that from the 4$^{\rm th}$ monitoring (AM-5 through AM-11 in Fig.~\ref{fig:decrease}), with a model function that considers physical  and environmental attenuation,
\begin{equation}\label{eqfit}
f(t)=N \left\{a_{\rm fast} 2^{-t/T_{\rm fast}}+ (1-a_{\rm fast})2^{-t/T_{\rm slow}}\right\}\cdot \frac{(k \times 2^{-t/T_{134}}+2^{-t/T_{137}})}{k+1}.
\end{equation}
Here, $T_{134}=2.06\,\rm y$ and $T_{137} = 30.17\,\rm y$  are, respectively, the physical half lives of $^{134}$Cs and  $^{137}$Cs, $T_{\rm fast} (=0.43\pm0.01 \rm \, y)$ and $T_{\rm slow} (> 400 \rm \, y)$ are the fast and slow half lives of environmental decay, and $a_{\rm fast} (= 0.60\pm 0.01)$ is the fraction of the fast decaying component, and $k=2.95$ is the ratio of air-kerma-rate constant of $^{134}$Cs to $^{137}$Cs~\cite{miyazakihayano2}. 
In the fit, the pre-factor $N$ was chosen to set the value of the model function to unity at the timing of the 4$^{\rm th}$ monitoring ($t=0.65$ y).
 
For year 3, the function $f(t)$ (Eq.~\ref{eqfit}) was integrated from $t=2\, \rm y$ to $t=3\,\rm y$ (shaded area A in Fig.~\ref{fig:decrease}) and  the result was compared with the value at the reference date of the 8th airborne monitoring (red dot at $y=2.69$). The correction factor for year 3 thus obtained was $ F = 1.05 $. For year 5, the integration was done from $t=4\,\rm y$ to $t=5\,\rm y$ (shaded area B in Fig.~\ref{fig:decrease}), which yielded a factor $ F = 1.03 $. This procedure is graphically presented in Fig.~\ref{fig:decrease} by gray rectangles drawn on top of area A and on B. We used these factors to convert the  ambient dose equivalent rate at each grid point $\dot{H}^* (10)$ ($\mu$Sv/h) to the annual additional ambient dose equivalent  $H^* (10)$ (mSv), as, $H^* (10)=F\times \frac{24\times 365}{1000}\times \dot{H}^* (10)$. The annual additional  effective dose from external exposure  maps created  using the factor $c=0.15$ were interpolated to draw  1 mSv annual  isodose lines.

\section{RESULTS}
\begin{figure}
\includegraphics[width=0.6\textwidth]{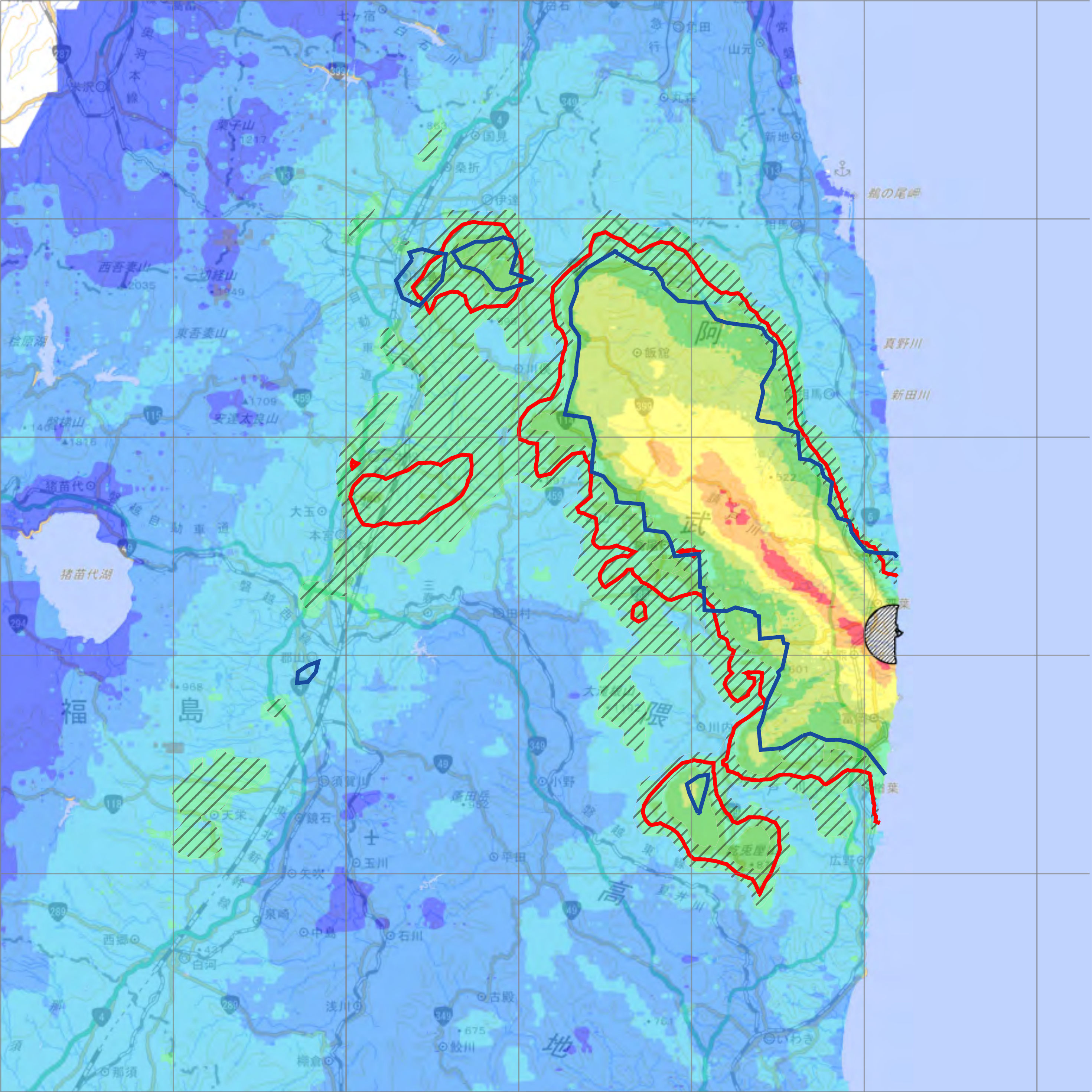}\includegraphics[width=0.2\textwidth]{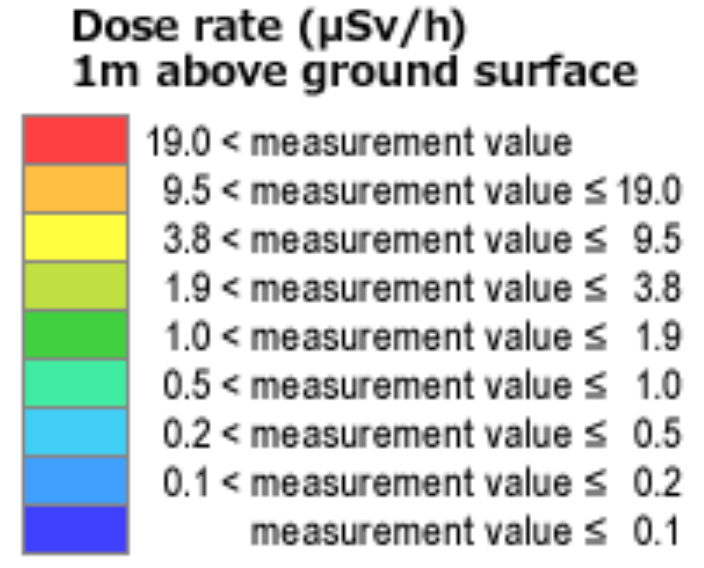}

\hspace*{5mm}

\includegraphics[width=0.6\textwidth]{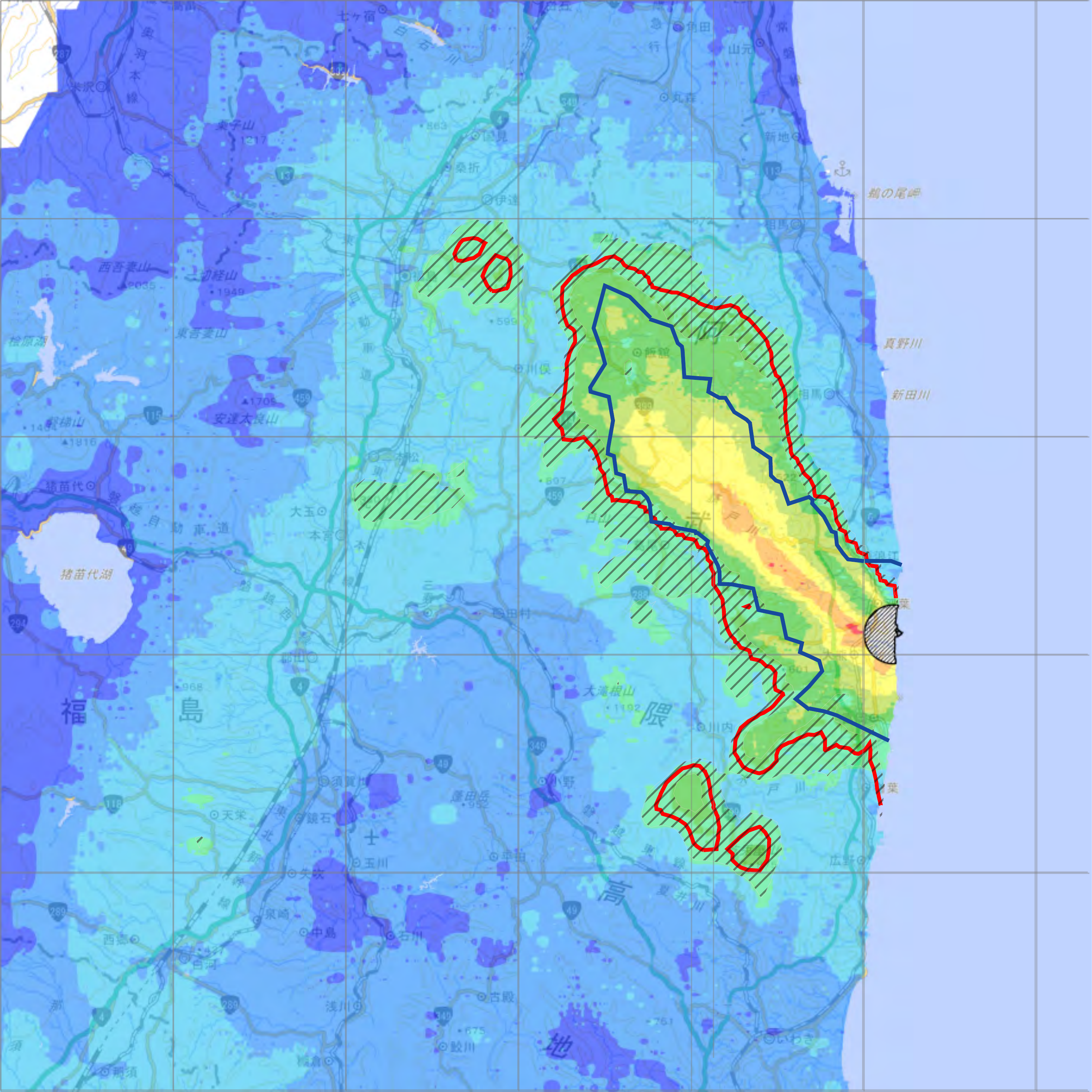}\includegraphics[width=0.2\textwidth]{hanrei.pdf}

\caption{\label{fig:map} 
Top: On the 8$^{\rm th}$ airborne monitoring map (reference date: November 19, 2013), annual 1 mSv isodose line for the third year as estimated by UNSCEAR (blue line) and that calculated by multiplying airborne monitoring result by c = 0.15  (red) are superimposed. Bottom: On the 10$^{\rm th}$ airborne monitoring map (reference date: November 2, 2015) annual 1 mSv isodose line (blue) for the fifth year of UNSCEAR and that calculated by multiplying airborne monitoring result by $c = 0.15$  (red) are superimposed.  In both panels, the hatched regions indicate the isodose bands between $c=0.10$ (25-percentile) and $c=0.22$ (75-percentile).}
\end{figure}

In Figs~\ref{fig:map} Top (Bottom), isodose lines for 
an annual additional  effective dose of 1 mSv 
 for 3 (5) years after the accident are overlaid on the 8$^{\rm th}$ (10$^{\rm th}$) airborne monitoring  map. The 1 mSv isodose lines shown in blue are the recalculated UNSCEAR estimate, and the contours in red are the 1 mSv isodose line estimated from the airborne monitoring results, converted to annual effective dose rates with the factor $c=0.15$. Also superimposed  are (hatched regions) the isodose bands between $c=0.10$ (25-percentile) and $c=0.22$ (75-percentile), reflecting the distribution of the factor $c$ as presented in Fig.~5 of Ref.~\cite{miyazakihayano1}. Presumably, uncertainties may be also inherent in the UNSCEAR's estimate (blue line), but they are not explicitly discussed in Ref.~\cite{unscear}.

In both years, the isodose lines  independently calculated with different methods are in fairly close agreement, but on closer examination, the UNSCEAR predicted isodose line (blue) encloses a smaller area than the line based on actual measurements derived in this study (red), the difference being slightly greater for the 5$^{\rm th}$ year.

\section{DISCUSSION}

After the FDNPP accident, an airborne monitoring method has been established and carried out regularly~\cite{sanada}, and the  ambient dose equivalent rates have been released as maps and numerical data~\cite{map}. 
Using the data of the large-scale individual dose monitoring conducted by Date City,  Fukushima Prefecture, Miyazaki et al.~\cite{miyazakihayano1} found that the personal dose equivalent ($\approx$ the effective dose) monitored by glass-badges, and the ambient dose equivalent of the residential area from airborne monitoring are closely correlated.

Naito et al.~\cite{naito}   used the airborne monitoring database and individual dosemeters (D-Shuttle) along with a global positioning system, possessed by approximately 100 voluntary participants, and obtained a conversion factor of   $c\sim 0.2$. In a more recent study by Naito et al.~\cite{naito2}, targeting 38 voluntary residents of Iitate village, Fukushima Prefecture, the coefficient was $c\sim 0.13$ for time spent at home and $c\sim 0.18$ for time spent outdoors. 

The results of these well-controlled studies targeting a small number of participants are thus consistent with the large-scale `glass-badge' result of Ref.~\cite{miyazakihayano1}. 
These works have established that the results of the airborne  monitoring can be used to assess the median  effective doses of the groups  residing in areas defined by the airborne-survey grid points. 
In the present study, we used $c=0.15$, the coefficient obtained from the Date-City glass-badge survey data.

UNSCEAR, in Att-2 of their 2015 White Paper~\cite{unscear}, predicted the long-term   cumulative effective dose ab extra after the Fukushima accident. In Att-2, estimates of  cumulative effective dose up to 100 years after the accident are shown in terms of  $^{137}$Cs deposition density as of June 14, 2011.   They also published the GIF animation published on the website at the same time, isodose lines for an annual additional effective dose ab extra in the 1$^{\rm st}$, 3$^{\rm rd}$, 5$^{\rm th}$, 10$^{\rm th}$, 30$^{\rm th}$ and 50$^{\rm th}$ years after the accident were presented. Comparison of the isodose line of annual additional   effective dose of 1 mSv estimated by UNSCEAR and the isodose line estimated from the data obtained by airborne monitoring with the factor $c = 0.15$~\cite{miyazakihayano1} shows relatively good agreement, despite the widely differing methods employed.

\begin{figure}
\includegraphics[width=0.9\textwidth]{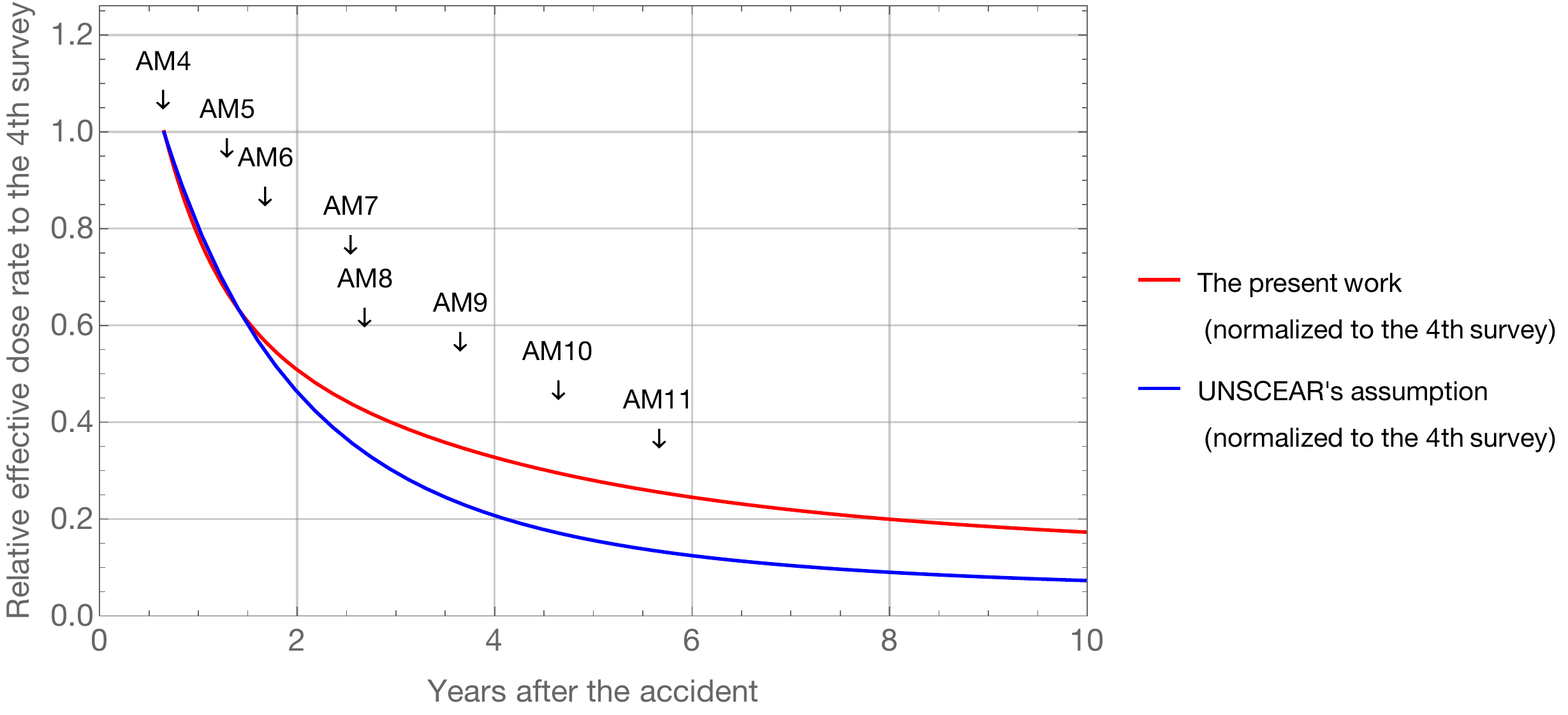}
\caption{\label{fig:comparison} The red curve is the time evolution of the median effective dose rate, relative to $t=0.65$ y (4$^{\rm th}$ airborne monitoring), the functional form of which is the same as the curve in Fig.~\ref{fig:decrease}. The blue curve is the time evolution of the effective dose rate calculated by using the UNSCEAR's assumptions (for details, see text).}
\end{figure}

However, on a closer examination, a trend in which UNSCEAR's isodose line encloses a smaller area than that generated in the present work is evident. The reason for this small but noticeable difference can be understood by comparing the attenuation of the ambient dose rate deduced from the airborne monitoring, with that assumed in the UNSCEAR's prediction.
The attenuation curve obtained by analyzing the airborne monitoring data  (in red), already shown in Fig~\ref{fig:decrease}, is reproduced in Fig.~\ref{fig:comparison}. In comparison, 
 a curve (in blue) for the attenuation of effective dose rates of representative population group calculated from the levels of deposition of radionuclides on the soil
is superimposed using the parameters assumed in the UNSCEAR's prediction, as described in detail in Ref.~\cite{attachmentc12}. In UNSCEAR's model, the attenuation $g(t)$ of the  effective dose rate  is factorized into three parts, 
\begin{equation}
g(t)={\rm physical}(t) \times {\rm environment}(t) \times {\rm location}(t).\label{eq:gt}
\end{equation}
Here, ${\rm physical}(t)$ is the physical decay curve of $^{134}$Cs and $^{137}$Cs, $ {\rm environment}(t)$ is the environmental attenuation with a 50-\% fast component half life of 1.5 y and a 50-\% slow component half life of 50 y, and ${\rm location}(t)$ is the location factor for typical adults (estimated to spend 0.6 of their time in wooden one-to-two-storey houses and 0.3 of their time at work in concrete multi-storey buildings), as described in Ref~\cite{attachmentc12}.

\begin{figure}
\includegraphics[width=0.9\textwidth]{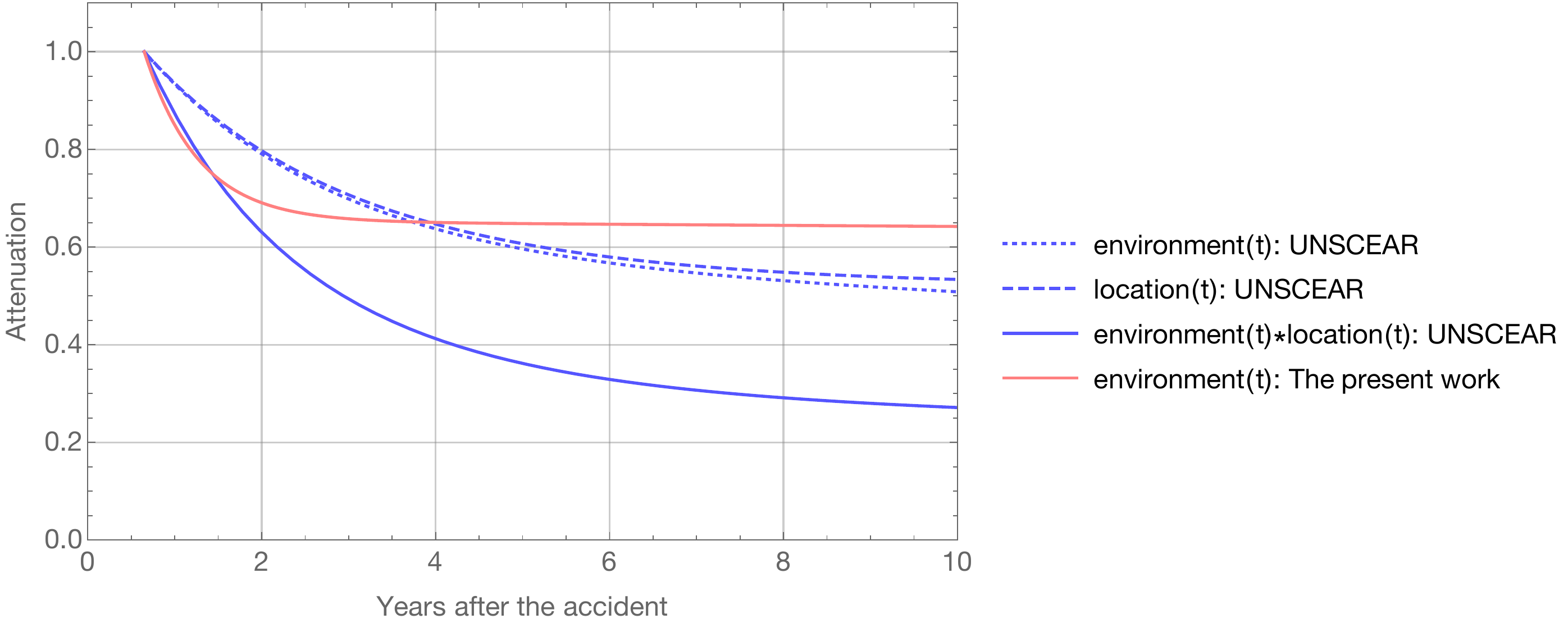}
\caption{\label{fig:breakdown} 
The graphs showing the contributions of ${\rm environment}(t)$ and ${\rm location}(t)$ to the time evolution of the effective dose rate. }
\end{figure}

The reason for the difference of two attenuation curves can be understood by showing environment(t) and local(t) of each curve individually (Fig.~\ref{fig:breakdown}). While Eq.~\ref{eqfit} uses smaller $T_{\rm fast}$ value and does not have ${\rm local}(t)$, which makes the curve a relatively rapid asymptote to the physical decay curve. The UNSCEAR's $g(t)$ has two, almost the similar magnitude of relatively slow attenuation functions, which together make greater reduction.  The authors deem both deposition density-to-dose conversion coefficient and ${\rm local}(t)$ based on the study of Chernobyl accident do not fit the Fukushima's case. 
Although UNSCEAR's prediction is useful for policy-makers, it is necessary to update based on the latest diachronic monitoring data continuously.

\section{Conclusions}

In this paper, we compared two different methods to estimate the individual external doses of the public residing in the Fukushima prefecture caused by radioactive fallout following FDNPP accident in 2011. One is the method adopted by UNSCEAR, which makes use of the official soil deposition map. The other makes use of airborne monitoring maps, together with an empirical factor $c=0.15$ obtained by comparing the large-scale `glass-badge' dosemeter measurements and the airborne surveys. Although the two are independent and are based on different data and methodologies, the resultant isodose lines for an annual additional  dose of 1 mSv  3 and 5 years after the FDNPP accidents show relatively good agreement.  However, to improve the accuracy of long-term annual effective  isodose lines, feedback from continuous measurements such as airborne monitoring is important.

\begin{acknowledgments}

The authors are grateful to Dr. J. Tada of Radiation Safety Forum for valuable discussions. This work was partially supported by donations by many individuals to RH through The University of Tokyo Foundation.

\end{acknowledgments}

\end{document}